\documentclass{jpp}
\usepackage{graphicx}
\usepackage{xcolor}

\usepackage[utf8]{inputenc}
\usepackage[T1]{fontenc}
\usepackage{amsmath}
\usepackage{bm}
\usepackage{float}

\usepackage{subcaption}
\usepackage{cleveref}
\usepackage{svg}

\shorttitle{Strain Optimization for ReBCO HTS Stellarator Coils in SIMSOPT}
\shortauthor{P. Huslage, et al.}

\title{Strain Optimization for ReBCO High-Temperature Superconducting Stellarator Coils in \texttt{SIMSOPT}}

\author{Paul Huslage\aff{1}
  \corresp{\email{paul.huslage@ipp.mpg.de}}, Elizabeth J. Paul\aff{2}, Mohammed Haque\aff{2}, Pedro F. Gil\aff{1}, Nicolo Foppiani\aff{3}, Jason Smoniewski\aff{1} \and  Eve V. Stenson\aff{1}}

\affiliation{\aff{1}Max-Planck Institute for Plasma Physics,
 85748 Garching, Germany
\aff{2}Columbia University, New York, NY 10027, USA
\aff{3}Proxima Fusion, 81671 München}

\begin{document}

\maketitle

\begin{abstract}
This work provides an optimization mechanism to ensure the compatibility of ReBCO (Rare-earth Barium Copper Oxide) high-temperature superconducting (HTS) tapes with non-planar stellarator coils. ReBCO coils enable higher field strengths and operating temperatures for the magnet systems of future fusion reactors but are sensitive to strain due to their brittle, ceramic functional layer. 
We have implemented a metric to optimize strain on stellarator coils made from ReBCO superconductors into the stellarator optimization framework \texttt{SIMSOPT} and used it to design new stellarator coil configurations. 
To ensure structural integrity of coils wound with HTS tape, we introduce a penalty on binormal curvature and torsion along a coil. It can be used to optimize the orientation of the winding path for a given coil filament or to jointly optimize orientation and coil filament. 
We apply the strain optimization to three cases. For the EPOS (Electrons and Positrons in an Optimized Stellarator) design, we combine the strain penalty with an objective for quasisymmetry into a single-stage optimization; this enables us to find a configuration with excellent quasisymmetry at the smallest possible size compatible with the use of ReBCO tape. For CSX (Columbia Stellarator eXperiment), in addition to HTS strain, we add a penalty to prevent net tape rotation to ease the coil winding process. If the strain is calculated for a coil at reactor scale, we find a considerable variation of the binormal and torsional strain over the cross section of the large winding pack (0.5\,m x 0.5\,m). By exploiting the overall orientation of the winding pack as a degree of freedom, we can reduce binormal and torsional strains below limits for every ReBCO stack.
\end{abstract}
\section{Introduction}
Stellarators break the axisymmetry of the magnetic field to create a rotational transform without requiring a large plasma current. However, modern optimized stellarators utilize complicated non-planar coils to achieve particle confinement. These present a considerable engineering challenge due to their complexity and low tolerances (\cite{rise_experiences_2009}, \cite{strykowsky_engineering_2009}).

Rare-earth Barium Copper Oxide (ReBCO) superconductors offer promising characteristics for the high-field magnets required in future fusion reactors. Unlike conventional low-temperature superconductors such as NbTi and Nb$_3$Sn, ReBCO exhibits enhanced tolerance for high magnetic field strength and can operate at higher temperatures, around 20\,K, as opposed to 4\,K, enabling more efficient cooling of the coils and a smaller device size. These superconductors are typically deposited as thin ($\approx 4\,\mu$m) ceramic layers onto Hastelloy bands or "tapes" (100\,$\mu$m thick and 3 mm to 12 mm wide) to render them usable. 

However, the brittleness of the ceramic layers still poses a challenge, particularly in the context of the intricate, non-planar configurations of modular stellarator coils. Addressing this challenge involves reducing strain to ensure compatibility with ReBCO by adjusting tape orientation along the coil. 

\begin{figure}
    \centering
    \begin{subfigure}[b]{1.0\textwidth}
        \centering
        \includegraphics[width=\textwidth]{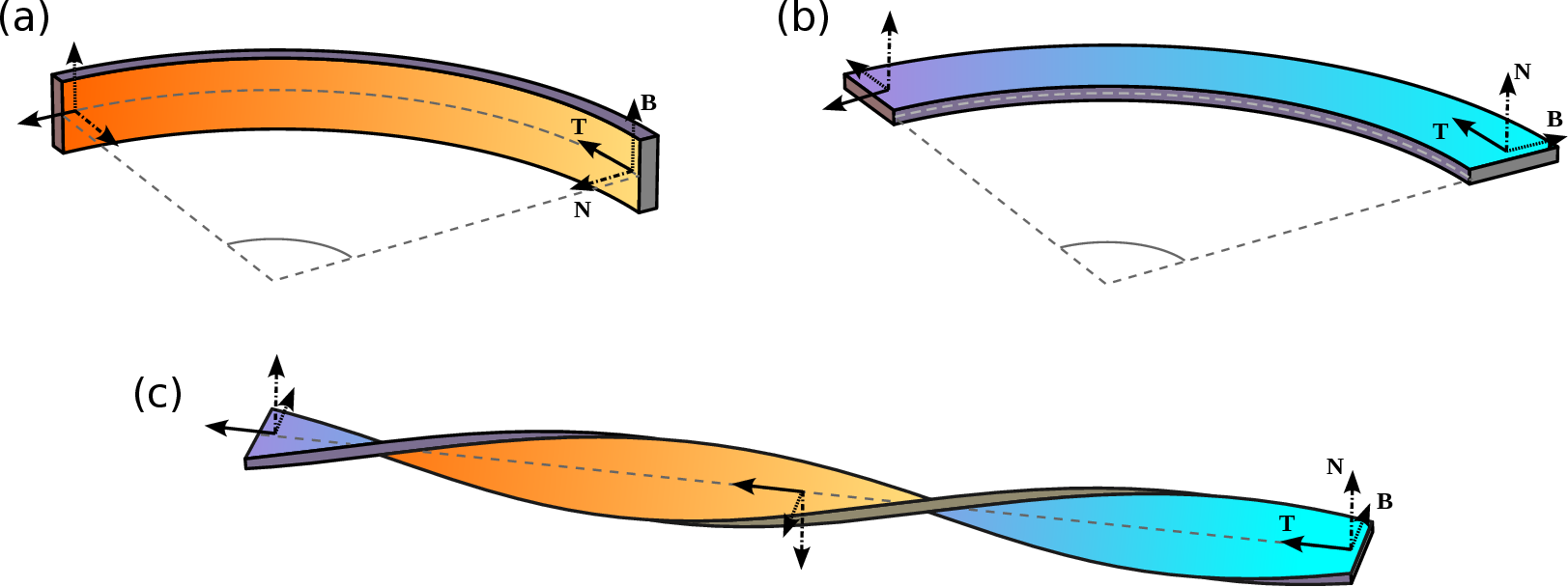}
    \end{subfigure}
    \hfill
    \caption{Three different ways to deform a superconducting tape. Normal (a) and binormal (b) curvature as well as torsion (c).}
    \label{fig:strains}
\end{figure}

There are three ways to apply strain to a tape, as illustrated in Fig. \ref{fig:strains}. The quasi two-dimensional nature results in a preferred direction of bending. Bending the tape around its normal axis ("hard-way bending" or "binormal curvature") strains the ReBCO layer more than compared to bending around the binormal axis ("easy-way bending" or "normal curvature"). In addition to the two forms of bending, the tape is also put under strain by twisting - a rotation of the tape frame ("torsion").
We assume a critical strain of 0.2\% - 0.4\% throughout this paper. While \cite{nickel_impact_2021} found no significant degradation of the critical current below 0.4\% strain, the EPOS and CSX projects choose a more conservative 0.2\% strain as the acceptable limit. This provides a safety margin for the assembly process as the strain during winding may exceed that of the finished coil. (E.g. a handling error led to a defect in one of the coils built in \cite{huslage_winding_2024})
Optimization of tape winding path orientation involves trading off hard-way bending and torsion to distribute these along a coil, enabling all parts of the coil to stay within strain limits.
Previous efforts by \cite{paz-soldan_non-planar_2020} and \cite{huslage_winding_2024} optimized the orientation of a tape stack wound into the shape of already existing coils.  Now, it is possible to include the strain penalty into the filamentary coil optimization process. It is not always possible to find a ReBCO-compatible winding path for a given filamentary coil shape. In those cases, it is important to incorporate the strain metric into the filamentary coil optimization to arrive at coils can actually be wound without damage to the conductor. 

This paper is structured as follows: Section \ref{sec:strain} introduces the metrics to optimize for strain on non-planar, 
ReBCO coils, along with a metric for minimizing full rotations of the tape frame used in CSX coil optimization. Section \ref{sec:implementation} discusses implementation into the stellarator optimization framework \texttt{SIMSOPT} \citep{landreman_simsopt_2021} and shows benchmarks againist previous methodologies. Sections \ref{sec:epos} and \ref{sec:csx} show strain-optimized coil sets for the small stellarators EPOS and CSX, respectively.   
Section \ref{sec:reactor} shows the application of the strain optimization on a reactor-relevant winding pack. Section \ref{sec:summary} summarises the results of this paper and gives an outlook for future work. 
\section{Metrics for strain on superconducting ReBCO coils}
\label{sec:strain}
In stellarator optimization, coils are often represented as filamentary curves (e.g. \cite{zhu_designing_2018}). The shape of these curves is typically optimized to fit the magnetic field on the boundary surface of a given magnetohydrodynamic equilibrium. 

In order to define an optimized path for the ReBCO tape, we need to add the orientation of the tape winding to the filamentary representation. We do this by defining a local frame at each point on the curve. In \texttt{SIMSOPT}, there are two such frames already implemented.

\begin{figure}
    \centering
    \includegraphics[width=0.9\textwidth]{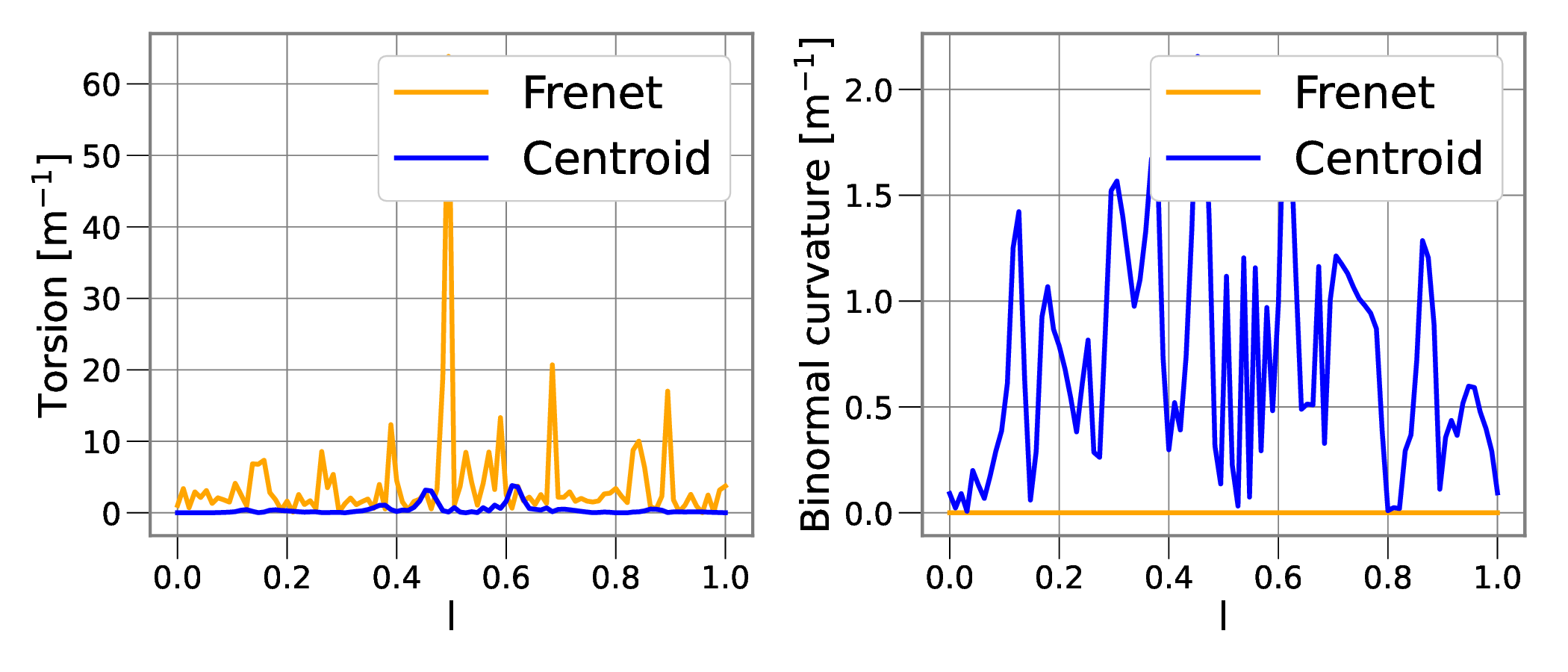}
    \caption{Torsion (left) and binormal curvature (right) for the Frenet (orange) and the centroid (blue) frames along a W7-X coil. The Frenet frame has no binormal curvature but high, localized torsion. The centroid frame has more regular  torsion but has strong binormal curvature. Both frames require optimization to be compatible with ReBCO.}
    \label{fig:frames}
\end{figure}

For a twice-differentiable, 3-dimensional curve $\boldsymbol{\Gamma}$, there exists a right-handed coordinate system called the Frenet frame. It consists of the tangent ($\mathbf T$), normal ($\mathbf N$) and binormal ($\mathbf B$) unit vectors which are defined as
\begin{align*}
   \mathbf T &= \frac{\boldsymbol \Gamma'}{\|\boldsymbol \Gamma'\|} \\
   \mathbf{N} &= \frac{\mathbf T'}{\|\mathbf T'\|}\\
   \mathbf{B} &= \mathbf T\times \mathbf N
\end{align*}
where the prime indicates the derivative with respect to the arc length. The normal vector of the Frenet frame tends to oscillate strongly if the coil is locally straight ($\mathbf T'\approx 0$, as in Fig \ref{fig:frames}). To avoid this behaviour \cite{singh_optimization_2020} implemented the centroid frame, where the normal vector is calculated as 
\begin{align}
    \mathbf N = \frac{\boldsymbol{\delta}-(\boldsymbol{\delta}\cdot\mathbf T)\mathbf T}{\|\boldsymbol{\delta}-(\boldsymbol{\delta}\cdot\mathbf T)\mathbf T\|},\quad\boldsymbol{\delta} = \boldsymbol{\Gamma} - \mathbf X_{c,0}.
\end{align}
This frame defines $\boldsymbol\delta$ as the vector from the center of the curve using $\boldsymbol{X}_{c,0}$, which is the zeroth order cosine Fourier coefficient of the curve). The normal vector is the normalized part of $\delta$ orthogonal to $\mathbf T$.

Having defined $(\mathbf T, \mathbf N, \mathbf B)$ as a frame to $\boldsymbol\Gamma$, we now define another frame rotated by an angle $\alpha$ in the $\mathbf{N}-\mathbf{B}$ plane:
\begin{align}
    \label{eq:rotated_frame}
    \Tilde{\mathbf T}&=\mathbf T \\\nonumber
    \Tilde{\mathbf N}&=\mathbf N \cos{\alpha} + \mathbf B \sin{\alpha}\\\nonumber
    \Tilde{\mathbf B}&=\mathbf B \cos{\alpha} - \mathbf N \sin{\alpha}. 
\end{align}
For the rotated frame, the normal ($\kappa$) and binormal ($\eta$) curvatures and the torsion ($\tau$) are expressed as dot products of the frame vectors and their derivatives along the curve
\begin{align*}
    \kappa&=\frac{\Tilde{\mathbf N} \cdot \Tilde{\mathbf T}^\prime}{\|\boldsymbol \Gamma'\|}\\
    \eta&=\frac{\Tilde{\mathbf B} \cdot \Tilde{\mathbf T}^\prime}{\|\boldsymbol \Gamma'\|}\\
    \tau&=\frac{\Tilde{\mathbf B} \cdot \Tilde{\mathbf N}^\prime}{\|\boldsymbol \Gamma'\|}.
\end{align*}
All quantities are normalized to $\|\boldsymbol \Gamma'\|$ to account for uneven spacing between quadrature points. Estimating a critical strain $\epsilon_{\text{crit}}$ on a superconducting tape of width $w$ and thickness $h$, the upper bounds for $\kappa, \eta, \tau$ are
\begin{align*}
    \kappa_{\text{crit}}&=\frac{2\epsilon_{\text{crit}}}{h}\\
    \eta_{\text{crit}}&=\frac{2\epsilon_{\text{crit}}}{w}\\
    \tau_{\text{crit}}&=\frac{\sqrt{12\epsilon_{\text{crit}}}}{w}.
\end{align*}
We can safely ignore the strain from normal bending as $\kappa_{\text{crit}}/\eta_{\text{crit}}=w/h\ll 1$. Typical values are $h\approx100\,\mu\text{m}, w=3\,\text{mm}-12\,\text{mm}$, and $\epsilon_{\text{crit}}=0.2\%-0.4\%$. We calculate the torsional ($\epsilon_{\text{tor}}=\tau^2 w^2/12$) and hard-way bending ($\epsilon_{\text{bend}}=w\eta/2$) strains and implement LP norm penalty functions for optimization
\begin{align}
\label{eq:strain metric}
    J_{\text{tor}/\text{bend}}=\frac{1}{p}\int_{\boldsymbol\Gamma}dl\,\text{max}(\epsilon_{\text{tor/bend}} - \epsilon_{\text{crit}, \text{tor}/\text{bend}}, 0)^p.
\end{align}
Equation \ref{eq:strain metric} gives a smooth metric to compute the relevant strains along a coil. The parameter $p$ (2 by default) can be chosen to more strongly penalize high peaks in the metric. The parameters $\epsilon_{\text{crit}, \text{tor}/\text{bend}}$ are thresholds above which strain is penalized by the objective functions. 

In addition to the strain, we penalize the angle of rotation between the tape frame and centroid frame, defined as $\alpha$ in \eqref{eq:rotated_frame} and denoted here by $\alpha_{\text{centroid}}$. This is motivated by the desire to reduce the net winding of the tape frame, which complicates the winding process. 
An LP norm penalty function is defined as
\begin{align}
\label{eq:j_twist}
    J_{\text{twist}} = \left(\frac{\int_{\Gamma} dl \, \alpha_{\text{centroid}}^p}{\int_{\Gamma} dl }  \right)^{1/p}. 
\end{align}
Note that the implementation of this objective enables $\alpha_{\text{centroid}}$ to be evaluated in any frame, not necessarily one defined with respect to the centroid frame. 

\section{Implementation into \texttt{SIMSOPT}}
\label{sec:implementation}
\texttt{SIMSOPT} is a framework for stellarator optimization based on \texttt{python} and \texttt{C++}. Among its capabilities are the ability to define geometric quantities, such as boundary surfaces of stellarator equilibria, filamentary coil shapes and the HTS tape orientation. These geometric quantities can be optimized using standard routines from \texttt{scipy}. An overview over the code can be found in \cite{landreman_simsopt_2021}. 

Coils are implemented as one-dimensional current-carrying filaments. In order to incorporate the rotation of the winding pack, we implemented the class \texttt{FramedCurve} that adds an angle to the filamentary coil that represents the rotation of a given initial frame in the $\mathbf{N}-\mathbf{B}$ plane as described in \eqref{eq:rotated_frame}. The curve as well as the angle is described by a Fourier series. The metrics for strain from torsion and binormal curvature defined in \eqref{eq:strain metric} are implemented as \texttt{Optimizable} objects that can be used in a penalty function for coil optimization. 

\begin{figure}
	\centering
    \includegraphics[width=0.7\textwidth]{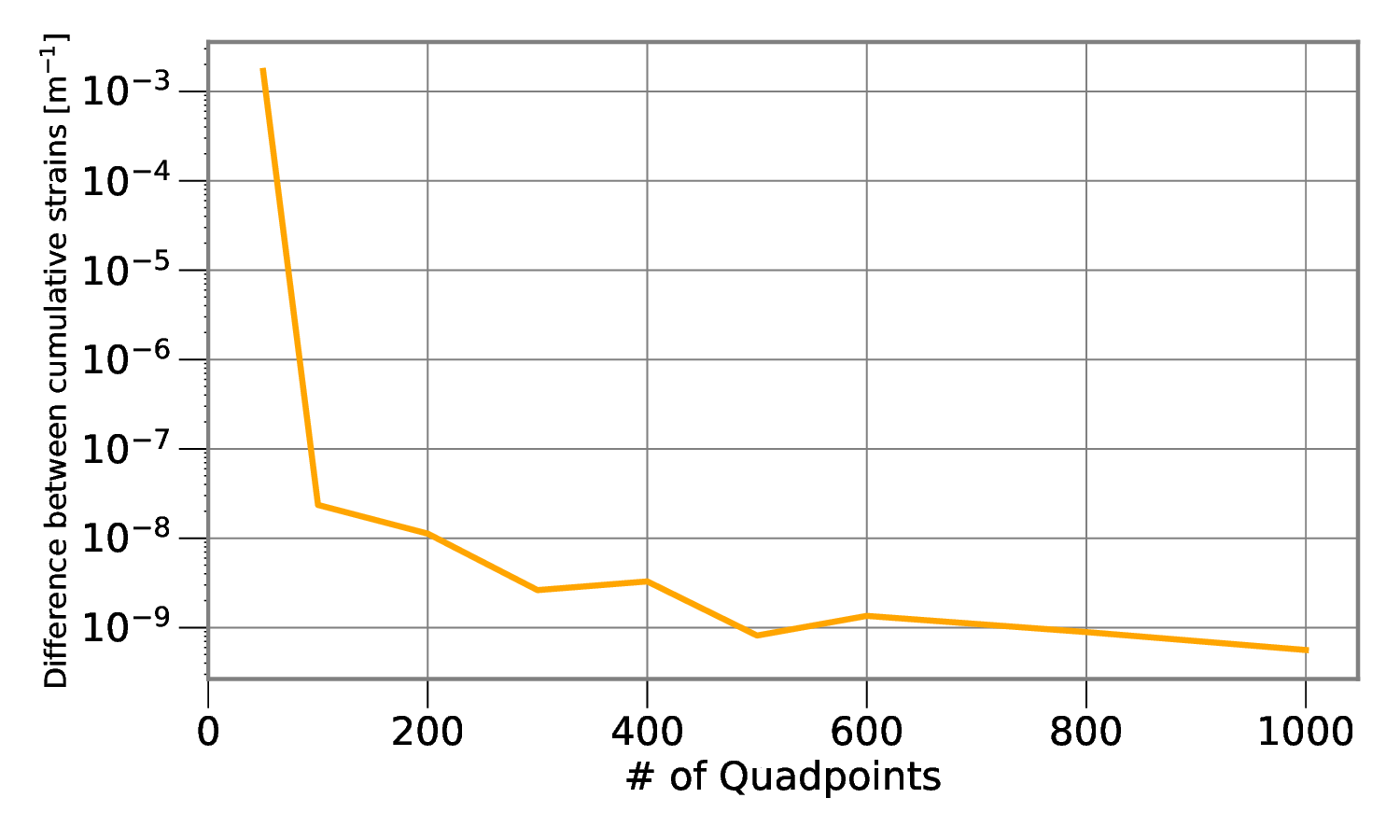}
	\caption{Cumulative torsional and hard-way bending strains from \eqref{eq:strain metric} for a model coil with different numbers of quadrature points. The difference $\Delta J_{strain}$ between strain penalties for subsequent numbers of quadrature points drops below single digit machine precision.}
    \label{fig:convergence}
\end{figure}

Figure \ref{fig:convergence} shows the values of the strain metric for a simple, non-planar model coil. The shape of the coil is kept fixed and only the tape orientation is changed. $J_{\text{strain}} = J_{\text{tor}} + J_{\text{bend}}$ has been evaluated with different numbers of quadrature points. The strain metric $J_{\text{strain}}$ is then averaged over fixed arc length intervals that are sampled with different numbers of quadrature points. Figure \ref{fig:convergence} shows the differences between the cumulative strain function for subsequent numbers of quadrature points, which drops below single digit machine precision. 

\begin{figure}
    \centering
    \includegraphics[width=1\linewidth]{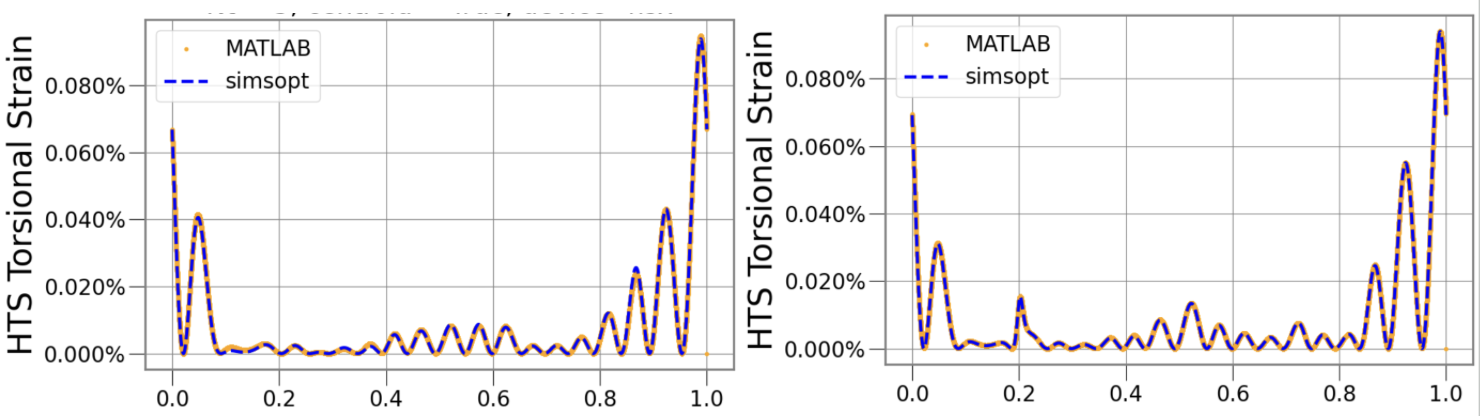}
    \caption{A comparison of HTS strain values for HSX coil \# 1, calculated with \texttt{SIMSOPT} and MATLAB implementation \citep{paz-soldan_non-planar_2020} in the centroid (left) and Frenet (right) frames.}
    \label{fig:comparison_matlab_simsopt}
\end{figure}

We are free to to fix the shape of the curve and optimize only the winding angle or include all degrees of freedom in the optimization (curve and angle). The first case does not change the curve itself but only optimizes the orientation of the ReBCO tape along the coil. This is similar to the approaches taken in \cite{paz-soldan_non-planar_2020} and \cite{huslage_winding_2024}. Figure \ref{fig:comparison_matlab_simsopt} shows a benchmark between the winding angle optimization used in \cite{paz-soldan_non-planar_2020} and our \texttt{SIMSOPT} implementation. The calculations are in excellent agreement with each other. 

However, not all filamentary coils are consistent with meeting strain requirements at the desired size, making it is necessary to include strain in the filamentary optimization. If we include all degrees of freedom (DOFs) of the framed curve into a coil optimization function, the shape of the filamentary curve is varied in addition to the winding angle. This allows for better optimized coils with lower strain while achieving a given target field. By changing the weight on the strain penalty term, it is possible to make a trade-off between field accuracy and complexity of the coils. A high weight on the strain penalty results in increasingly simple coils at the cost of increased field error.

This is similar to the impact of the regularization term used in the coil optimization code \texttt{REGCOIL}. Here, a current potential on a winding surface is optimized to fit a given magnetic equilibrium. The complexity of this surface is regulated by the parameter $\lambda$. Details on this code can be found in \cite{landreman_improved_2017}.

\begin{figure}
    \centering
    \includegraphics[width=0.7\linewidth]{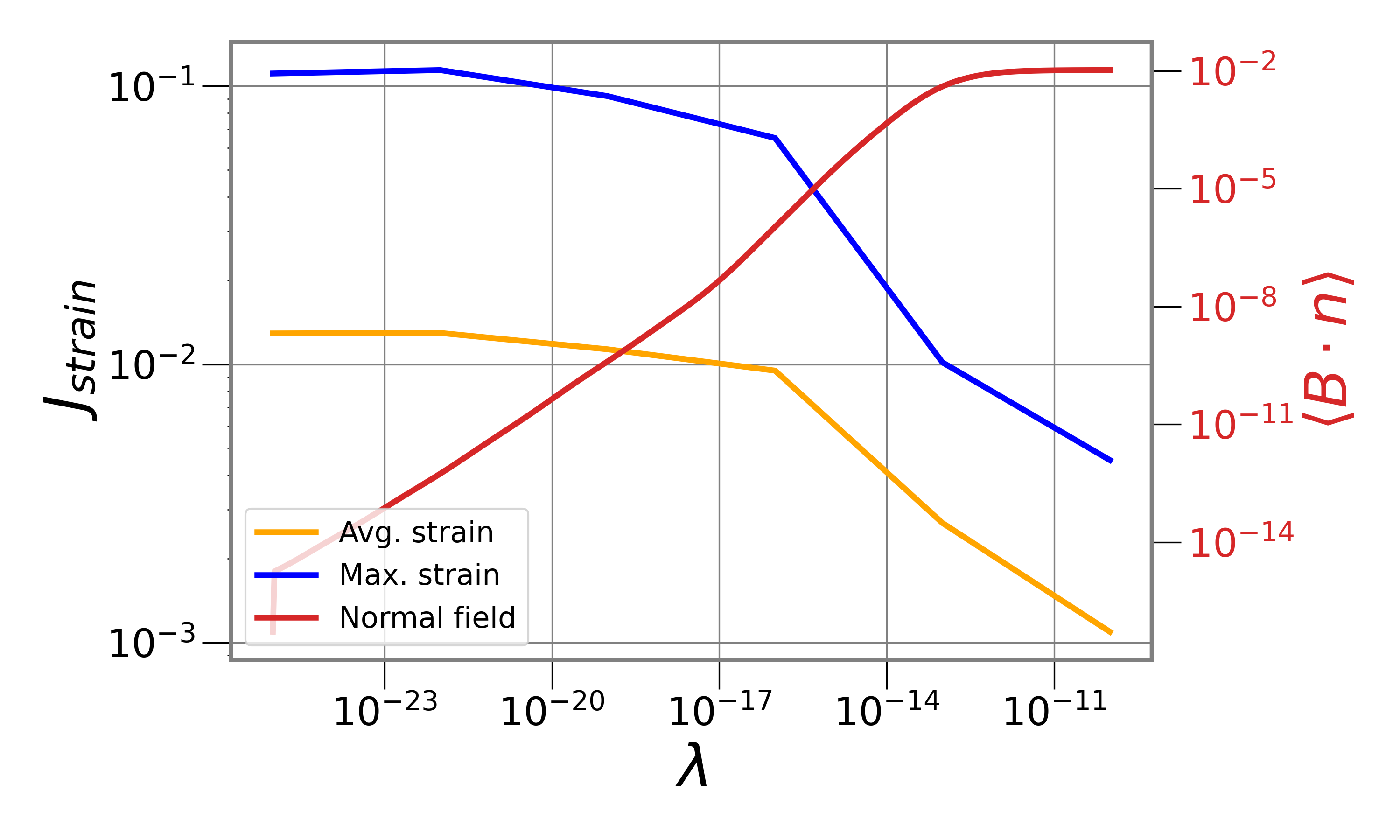}
    \caption{Average and peak strain penalty along coils created with \texttt{REGCOIL} code at different regularization compared to the corresponding averaged normal field. Increased coil regularization corresponds to lower strain along the coil, but also a higher field error.}
    \label{fig:comparison_regcoil_vs_strain}
\end{figure}

To compare the impacts of $\lambda$ and the strain penalty on the coil geometry, we generate coils for an EPOS candidate equilibrium in \texttt{REGCOIL} with different regularization parameters. These coils are imported into \texttt{SIMSOPT}, and the strain penalty is calculated using the centroid frame. 
Figure \ref{fig:comparison_regcoil_vs_strain} shows the peak strain values on these coils for different regularization penalties. The strain on the coils decreases for higher regularization at the cost of a higher field error.

\section{EPOS}
\label{sec:epos}
The EPOS experiment (\cite{stoneking_new_2020}) aims to confine an electron-positron pair plasma in a tabletop-sized stellarator. Leading design candidates use modular, non-insulated coils made from ReBCO tape to create a quasiaxisymmetric 2\,T magnetic field. The coils will be wound on 3D printed support structures that provide orientation for the ReBCO tape stack to follow a strain-optimized path. To fuel positrons into the confining magnetic field via $\mathbf E \times \mathbf B$ drift injection (\cite{stenson_lossless_2018}), two larger weave lane coils are required. They create stray field to couple to the positron beam and guide the particles tangentially to the stellarator magnetic field. With a limited number of positrons available, EPOS needs to be sufficiently small to reach the plasma state.
The Debye length $\lambda_D$ must be much smaller that the minor radius of the device $a$. This ratio scales as $a/\lambda_D\sim R^{-1/2}$. Therefore, we aim to reduce the major radius as much as possible.

Due to its small size ($R\approx0.2$ m), strain optimization is crucial. The coils need to produce the desired magnetic field while not exceeding the tight limits on torsion and hard-way bending. To achieve this, we include all of the DOFs (curve and winding angle) into the penalty function; additionally, we are utilizing a single-stage approach adopted from \cite{jorge_single-stage_2023}. This method minimizes cost functions for the equilibrium and for the coils, coupled via the squared flux term. For this equilibrium, we optimize for quasiaxisymmetry, $\iota = 0.101$, and a smaller major radius. For the coils, we include terms for the minimum coil-to-coil and coil-to-surface distance in addition to the strain penalty and squared flux. The parameters in equation \ref{eq:strain metric} are chosen as $p=2$ and $\epsilon_{\text{crit}, \text{tor}/\text{bend}}=0$.

\begin{figure}
    \centering
    \includegraphics[width=0.7\textwidth]{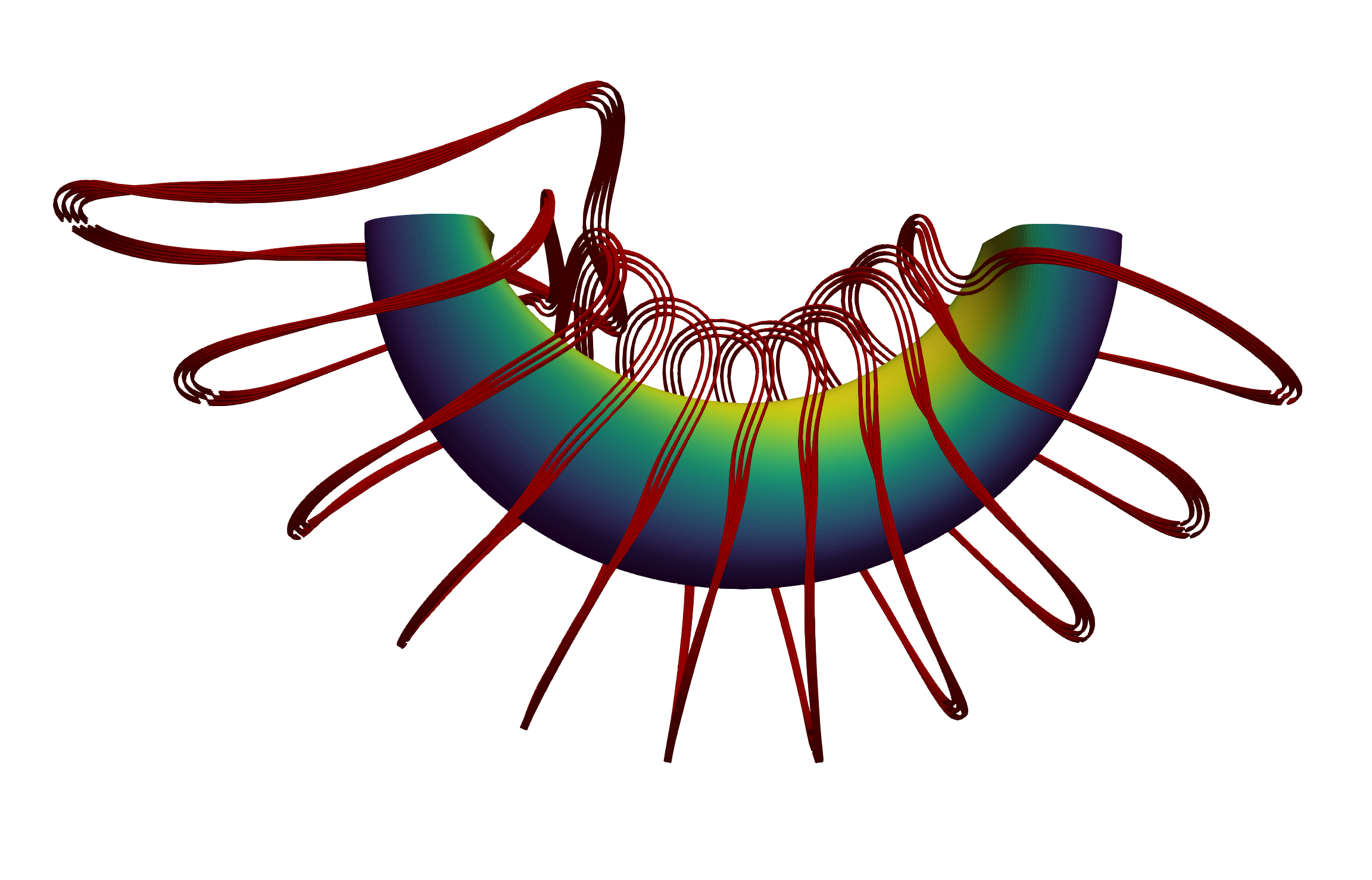}
    \caption{Optimized coil set within strain limits for a candidate equilibrium for EPOS. The coils are displayed in red. The yellow to blue color scale indicates the field strength on the outermost closed flux surface in the magnetic field created by the coil set.}
    \label{fig:epos_coils}
\end{figure}

\begin{figure}
    \centering
    \includegraphics[width=0.7\textwidth]{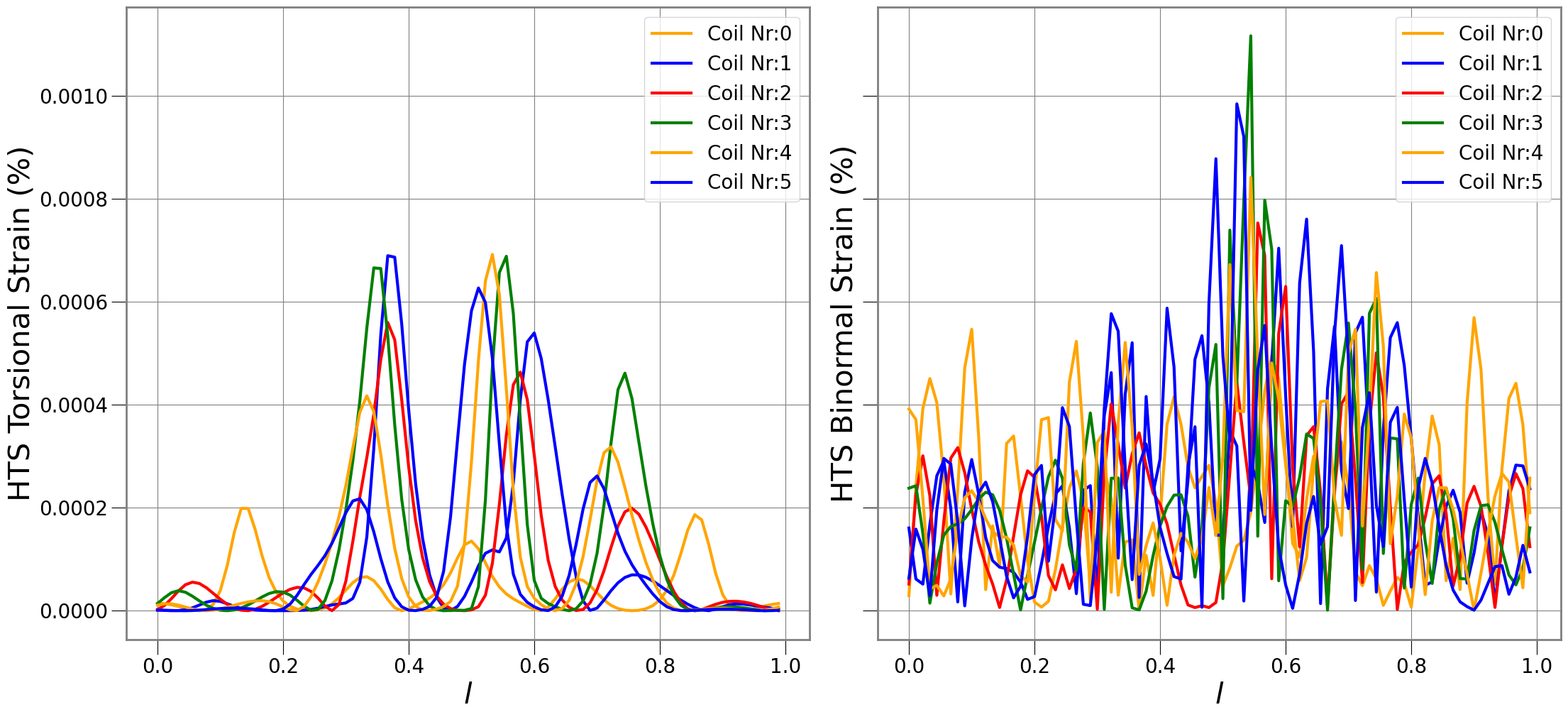}
     \caption{Torsional (left) and binormal curvature (right) strain for the coil set shown in figure \ref{fig:epos_coils}. None of the coils exceeds the strain limit of 0.2\% on 3\,mm wide ReBCO tape.} 
    \label{fig:epos_strains}
\end{figure}

\begin{figure}
    \centering
    \includegraphics[width=0.4\textwidth]{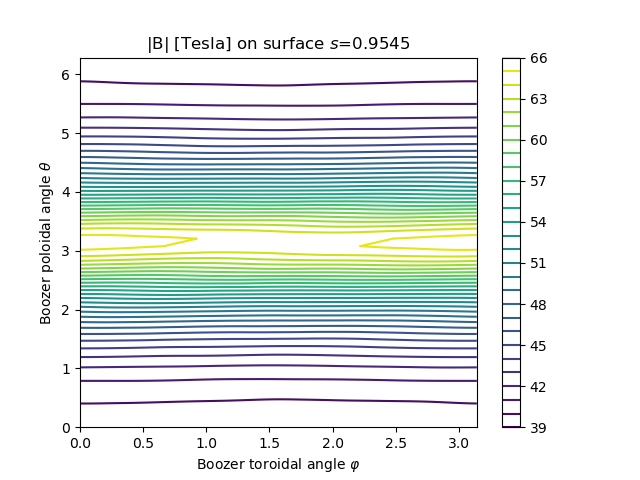}
    \includegraphics[width=0.4\textwidth]{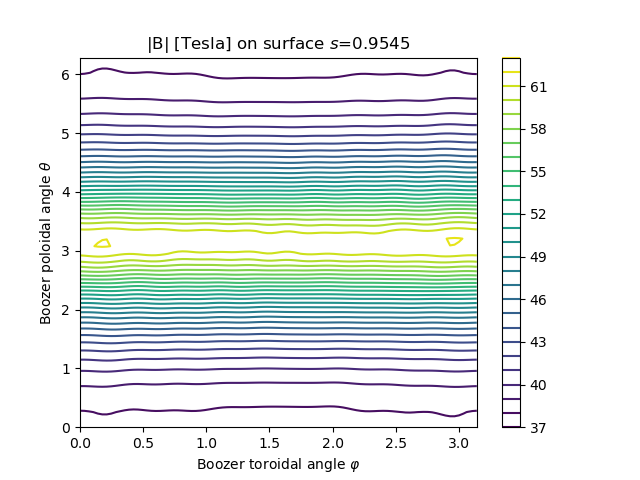}
    \caption{The field strength contours in Boozer coordinates for the EPOS candidate equilibrium (left) and the field produced by the strain optimized coils. (right). The quasisymmetry is only slightly degraded by the strain-optimized coils.}
    \label{fig:epos_error}
\end{figure}

The coil set displayed in figure \ref{fig:epos_coils} reproduces the magnetic field sufficiently well while not exceeding the strain limits on torsion and binormal curvature. Figure \ref{fig:epos_strains} shows the strain from torsion and binormal curvature along the coils. Figure \ref{fig:epos_error} displays the contours of the magnetic field strength in Boozer coordinates for the equilibrium and the magnetic field created from the coils. The coils retain good quasisymmetry and introduce only small symmetry-breaking modes. 

To improve the manufacturing and placement tolerances on the coil, we aim to combine the single-stage optimization with a stochastic approach. The full optimization cost function and the physics design of the EPOS stellarator will be detailed in an upcoming publication.
\newpage

\section{CSX}
\label{sec:csx}

The Columbia Non-neutral Torus (CNT) was designed to investigate non-neutral plasmas confined to magnetic surfaces using planar coils \citep{pedersen_construction_2006}. The magnetic configuration consists of two outer polodial field (PF) coils  and two interlocking (IL) coils within the vacuum vessel. The Columbia Stellarator eXperiment (CSX) will repurpose the PF coils and vacuum vessel of CNT to confine quasineutral plasmas with a quasiaxisymmetric magnetic field. The existing planar copper IL coils will be replaced by non-planar coils wound with non-insulated HTS tape on a 3D printed bobbin. In constructing this experiment, we aim to validate the physics of quasiaxisymmetry as well as demonstrate the compatability of ReBCO tape for stellarators. 

\begin{figure}
    \centering
    \includegraphics[width=0.5\textwidth]{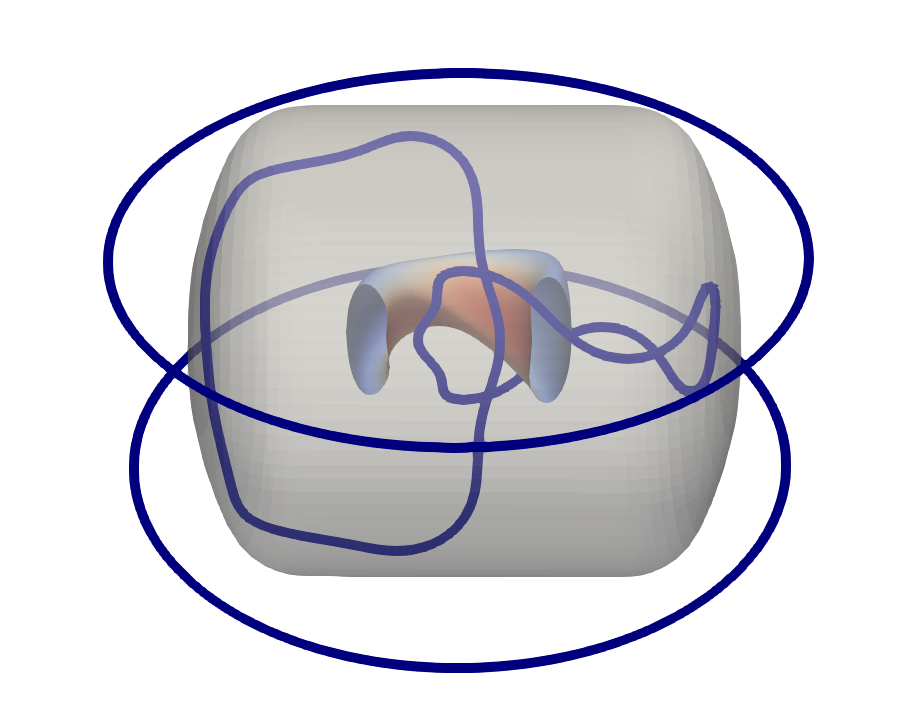}
    \caption{The preliminary CSX design consists of two planar PF coils outside the vessel and two non-planar interlinked coils inside the vessel, producing a quasiaxisymmetric equilibrium.}
    \label{fig:csx_vessel}
\end{figure}

\begin{figure}
    \centering
    \includegraphics[width=1\linewidth]{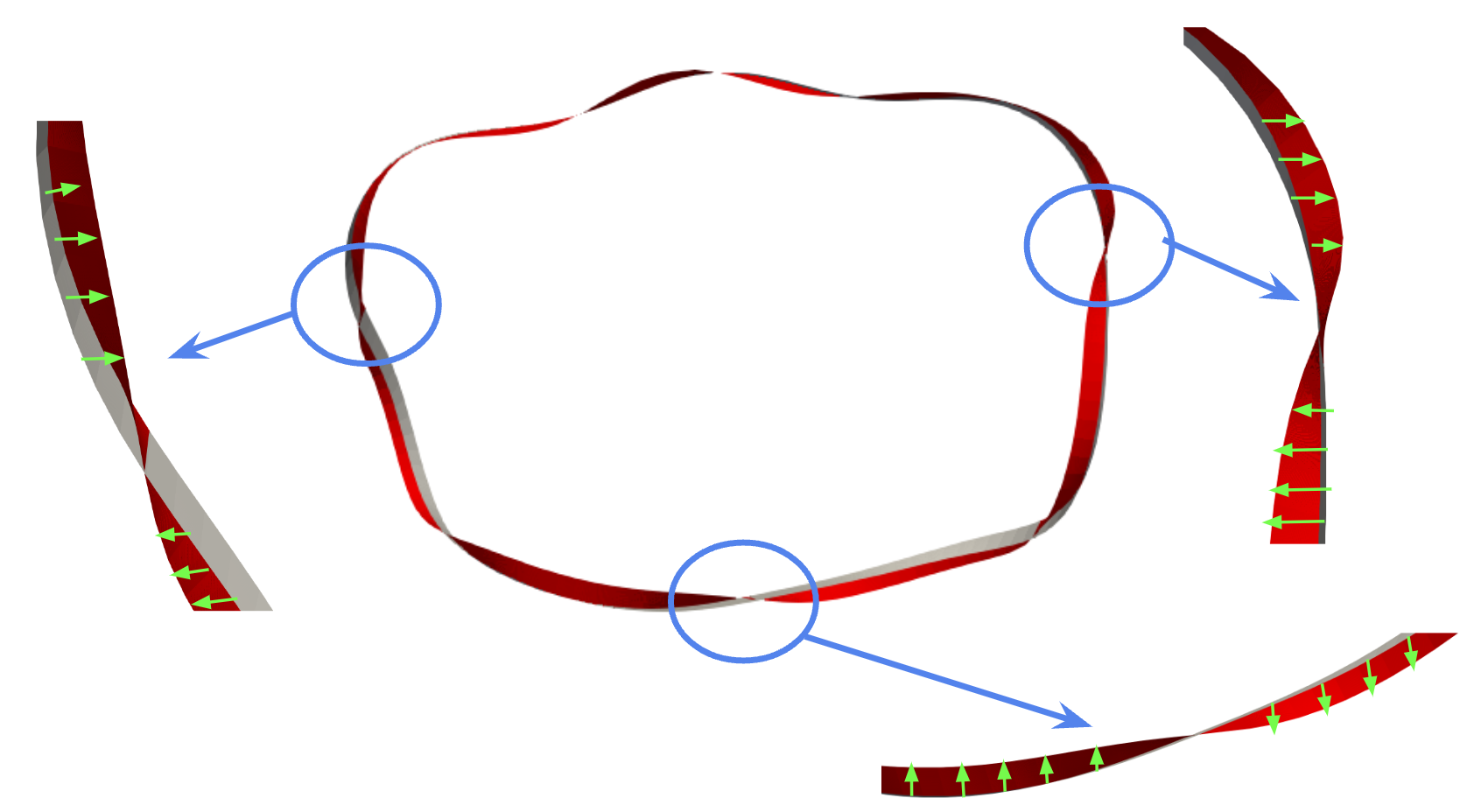}
    \caption{The tape orientation of the strain-optimized preliminary CSX coil is shown (red) in relation to the centroid frame (gray). The tape frame avoids any net rotations with respect to the centroid frame, as desired.}
    \label{fig:csx_frame}
\end{figure}

We present a preliminary IL coil design optimized for quasiaxisymmetry, $\iota \ge 0.15$, and plasma volume 0.1 m$^3$ (Fig. \ref{fig:csx_vessel} and \ref{fig:csx_frame}). Single-stage optimization is performed using the Boozer surface approach, in which approximate magnetic surfaces of the vacuum field are computed in Boozer coordinates with a least-squares solver \citep{giuliani_direct_2023}. The plasma surface, IL curve, and tape orientation DOFs are included in the optimization. Coil constraints are imposed to ensure sufficient clearance between the IL coils and the vacuum vessel, and the coil length is constrained to be $\le 4.25$ m. Binormal curvature and torsional strain are penalized using \eqref{eq:strain metric} with thresholds 0.2\% and a tape width of 4 mm. To ease tape winding, we include the twist penalty function $J_{\text{twist}}$ to avoid any net tape frame rotations. To facilitate winding under tension, we also penalize the rotation angle between the centroid and Frenet frames of the tape using \eqref{eq:j_twist} evaluated for the Frenet frame. This reduces the concave features visualized at the top of the coil in Figure \ref{fig:csx_frame}. The single-stage optimization result is shown in figure \ref{fig:csx_vessel}. 

\begin{figure}
    \centering
    \includegraphics[width=1\linewidth]{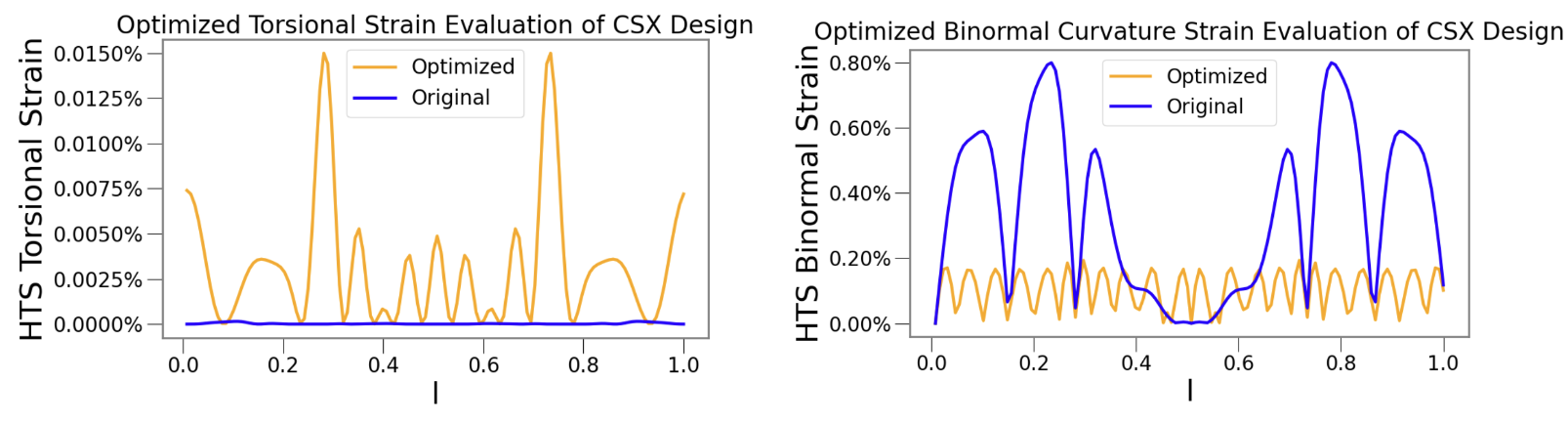}
    \caption{The torsional (left) and binormal curvature (right) HTS strain values are shown for the preliminary CSX coil. The curve obtained from single-stage optimization is fixed, and the winding path is optimized to obtain acceptable values of HTS strain.}
    \label{fig:csx_strains}
\end{figure}

After the single-stage optimization, the curve shape is fixed and the winding angle is again optimized for strain and frame rotation. The optimized tape path is shown in figure \ref{fig:csx_frame} (red), which does not deviate significantly from the centroid path (gray). The optimized strain values are below the chosen threshold, as seen in figure \ref{fig:csx_strains}. More details of the single-stage optimization will be presented in a future publication. 



\section{Reactor Winding Pack}
\label{sec:reactor}

In this section we demonstrate how strain optimization allows to mitigate curvature- and torsion-induced strains for an HTS coil at reactor scale.
Reactor-scale coils differ from smaller experiments as they require multiple stacks of tape to reach the desired current.
These currents for a reactor are in the range of 5-15 MA, and it would be impossible to realize them out of a single stack, as modern power supplies rarely achieve currents above 50 kA.
Such a coil requires 100-300 stacks, with total winding pack cross section sizes on the order of 10\% of the coil average radius.
This implies that the values of curvature and torsion can change by more than 10\% across stacks, which requires not only the optimization of the strain for a central filament, but for all stacks.
Note that, because stacks are wound in series, strain damage to any single portion of the HTS could make the entire coil fail.

For this study we focus on the coils of the Weldenstein 7-X experiment, scaled at the reactor design point proposed for the ARIES-CS power plant, defined by the minor radius of 1.7 m and the average field on axis of 5.7 T \citep{najmabadi_aries-cs_2008}.
We also assume that HTS will cover 16\% of the cross section of the winding pack, and we consider a HTS tape stack of 12 x 12 mm, as 12 mm is the most popular HTS tape width from manufacturing companies.
The winding pack cross section side length $s$ is determined by a simple algorithm that accounts for the available space, assuming that two coils are almost touching at the point where the coil filaments get the closest to each other:
\[ s = \frac{min(d_{cc})}{\sqrt{2}} - t, \]
where $d_{cc}$ is the distance between two coils, and the minium is taken across all coils, and $t$ is the case thickness, here assumed to be 5 cm.
The additional factor of $\sqrt{2}$ accounts for the diagonal of the square cross section.
This way the coil orientation can be chosen arbitrarily, preventing any possible clash between adjacent coils when choosing the orientation to perform strain optimization.

\begin{figure}
    \centering
    \includegraphics[width=0.48\linewidth]{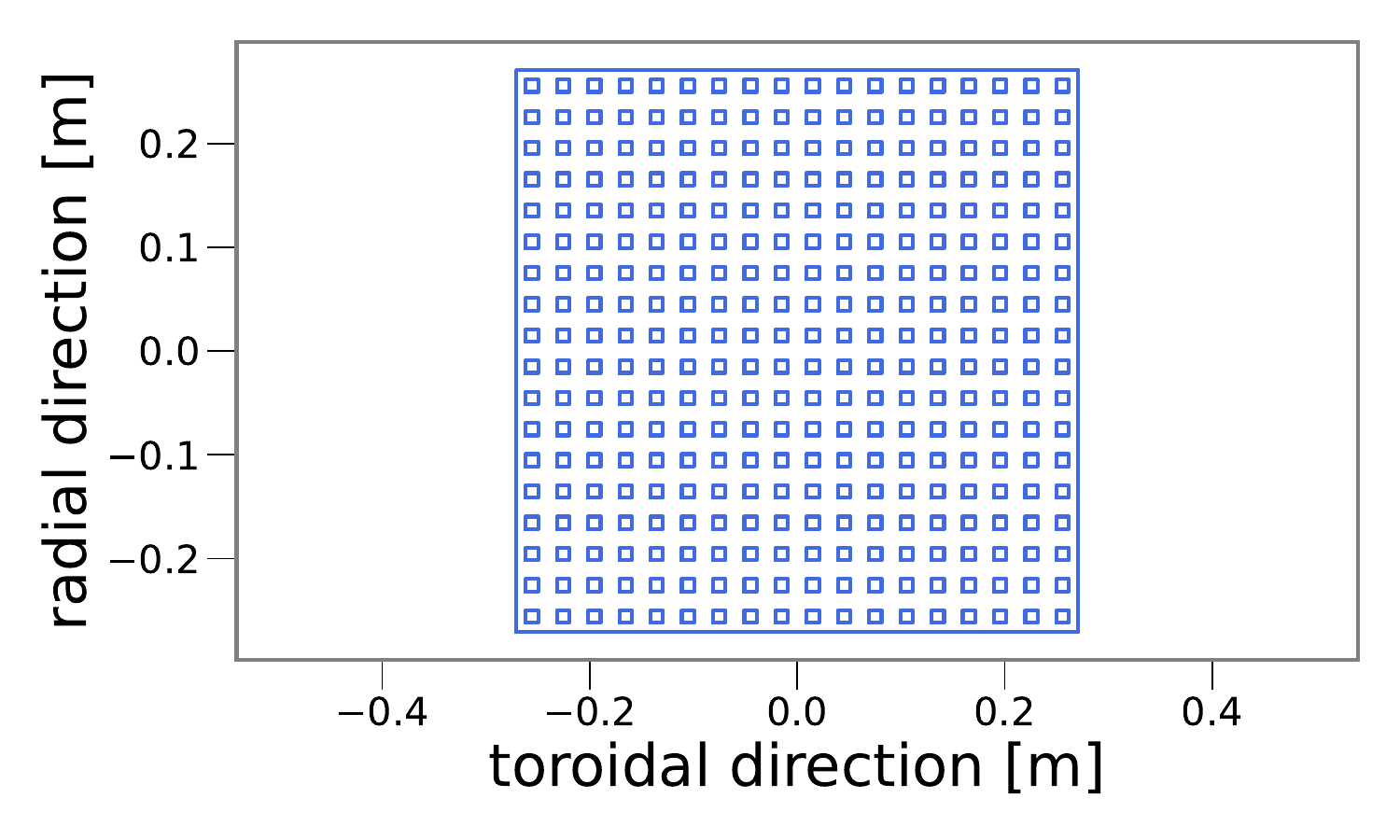}
    \includegraphics[width=0.48\linewidth]{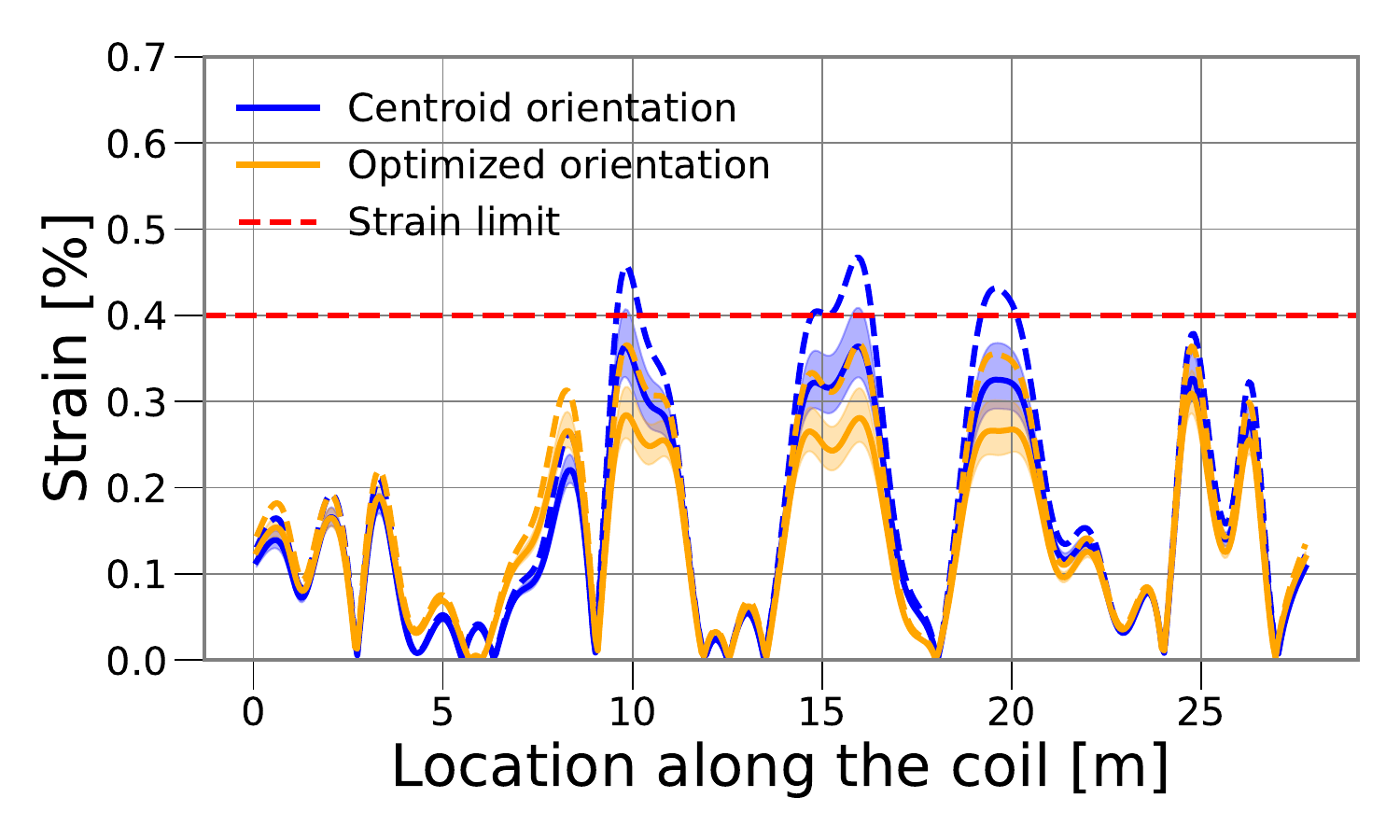}
    \caption{Left: cross section of a winding pack displaying the 324 ReBCO stacks.
    Right: strain profile along the first coil, comparing the pure centroid orientation (blue) with the optimized one (orange).
    The solid line represents the median value, the shaded band illustrates the region between the 16\% and the 84\% quantiles, and the dashed line the maximum across each stack.
    The red dashed line shows the 0.4\% strain limit.}
    \label{fig:reactor_cross_section_and_strain_before_after}
\end{figure}

With these design criteria, we obtained coils with a cross section of 54 x 54 cm, carrying a total current of 10.8 MA, 33.4 kA in each of the 324 stacks.
The cross section of the winding pack is schematically represented in the left plot of \cref{fig:reactor_cross_section_and_strain_before_after}.
All the relevant information is summarized in \cref{tab:reactor_summary_table}.

\begin{table}
\centering
\begin{tabular}{ll}
Minor radius             & 1.7 m   \\
Average field on axis    & 5.7 T   \\
Turn current             & 33.4 kA \\
Coil current             & 10.8 MA \\
Number of turns          & 324     \\
Winding pack side length & 54 cm   \\
Case thickness           & 5 cm    \\
Tape width           & 12 mm   \\
Tape stack size      & 12 mm   \\
Tape cross-section area fraction        & 16 \%
\end{tabular}
\caption{Reactor winding pack design parameters at the ARIES-CS design point \citep{najmabadi_aries-cs_2008}.}
\label{tab:reactor_summary_table}
\end{table}

We consider a cable realized as HTS tape stack first wound and later soldered.
In this case, the relevant strains are the torsional and binormal, as the tapes can freely slide on each other, reducing the strain in the normal direction.
We aim at reducing the strain below 0.4\% at each stack in each coil.

The strain of each stack depends on its local orientation, which can be modified by changing either the orientation of the entire winding pack, or the orientation of each stack individually.
Allowing for a free orientation of each stack is a significant challenge from an engineering perspective.
Another possibility is to modify the winding pack shape from a square cross section with stacks arranged on a regular grid to a different shape.
This option would however require more engineering design and it is therefore, outside of the scope of this paper.
Therefore, we limit ourselves to the optimization of the winding pack orientation.

In order to not modify the plasma equilibrium, we fix the centroid position, and we optimize for the orientation of the winding pack only.
The right plot of figure \ref{fig:reactor_cross_section_and_strain_before_after} shows the strain profile along the first of the five independent coils in the coilset, for the pure centroid orientation (orange) and the optimized orientation (blue).
The solid line shows the average strain, while the shaded region covers the region of one standard deviation around the mean value.
The dashed line illustrates the maximum strain.
In this case, the maximum exceeds the 0.4\% threshold in multiple regions.
The optimized configuration lies safely below the limits at each point.

\begin{figure}
    \centering
    \includegraphics[width=0.48\linewidth]{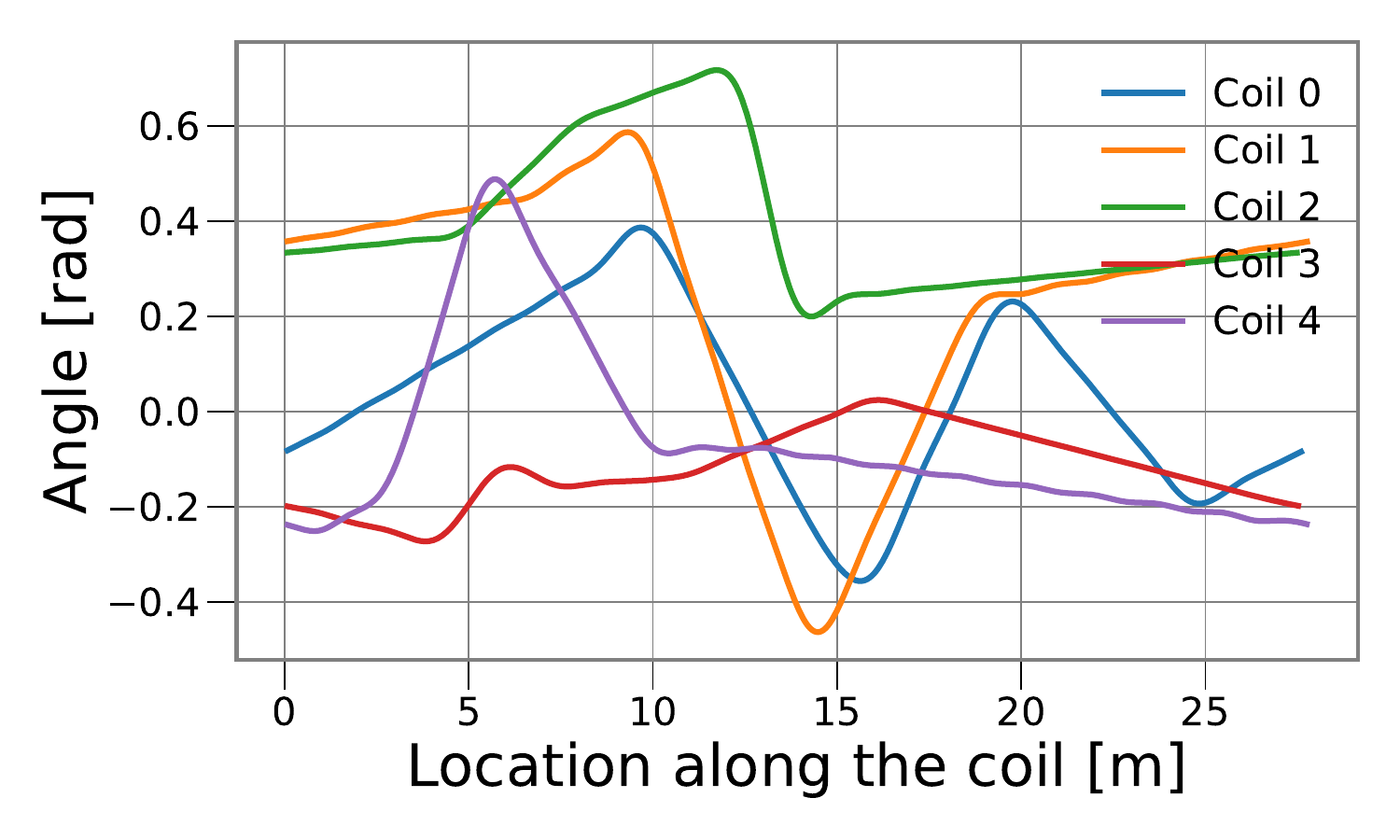}
    \includegraphics[width=0.48\linewidth]{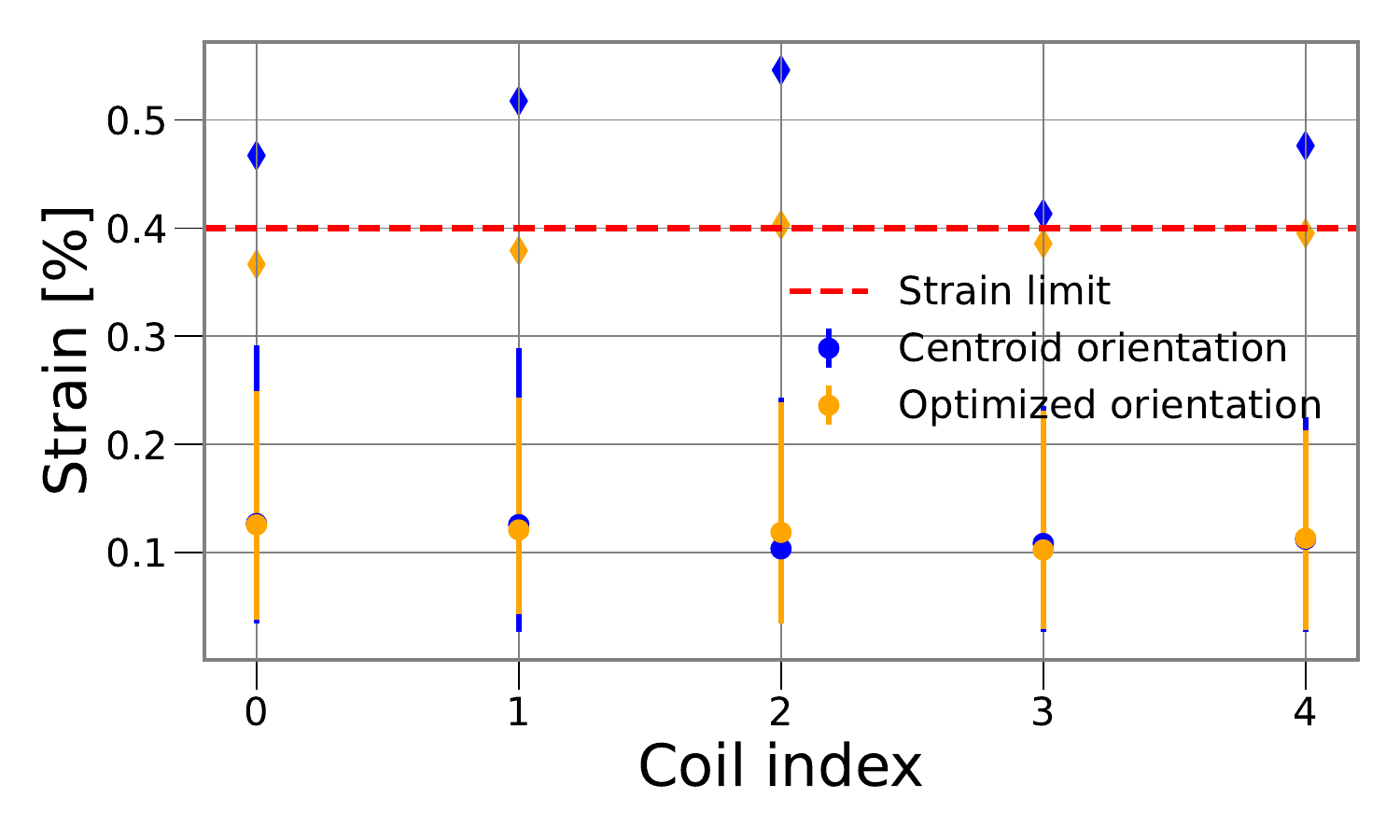}
    \caption{Left: the rotation angle as a function of the location of the coil, to be applied to the centroid frame to realize the strain-optimized orientation, for each coil in the coilset.
    Right: for each independent coil in the coilset (x-axis), the strain profile is summarized by the mean (solid circle), the standard deviation (error bar), and the maximum (diamond) for pure centroid orientation (blue) and the optimized one (orange).
    The red dashed line shows the 0.4\% strain limit.}
    \label{fig:reactor_rotation_angles_and_strain_max}
\end{figure}

\begin{figure}
    \centering
    \includegraphics[width=0.95\linewidth]{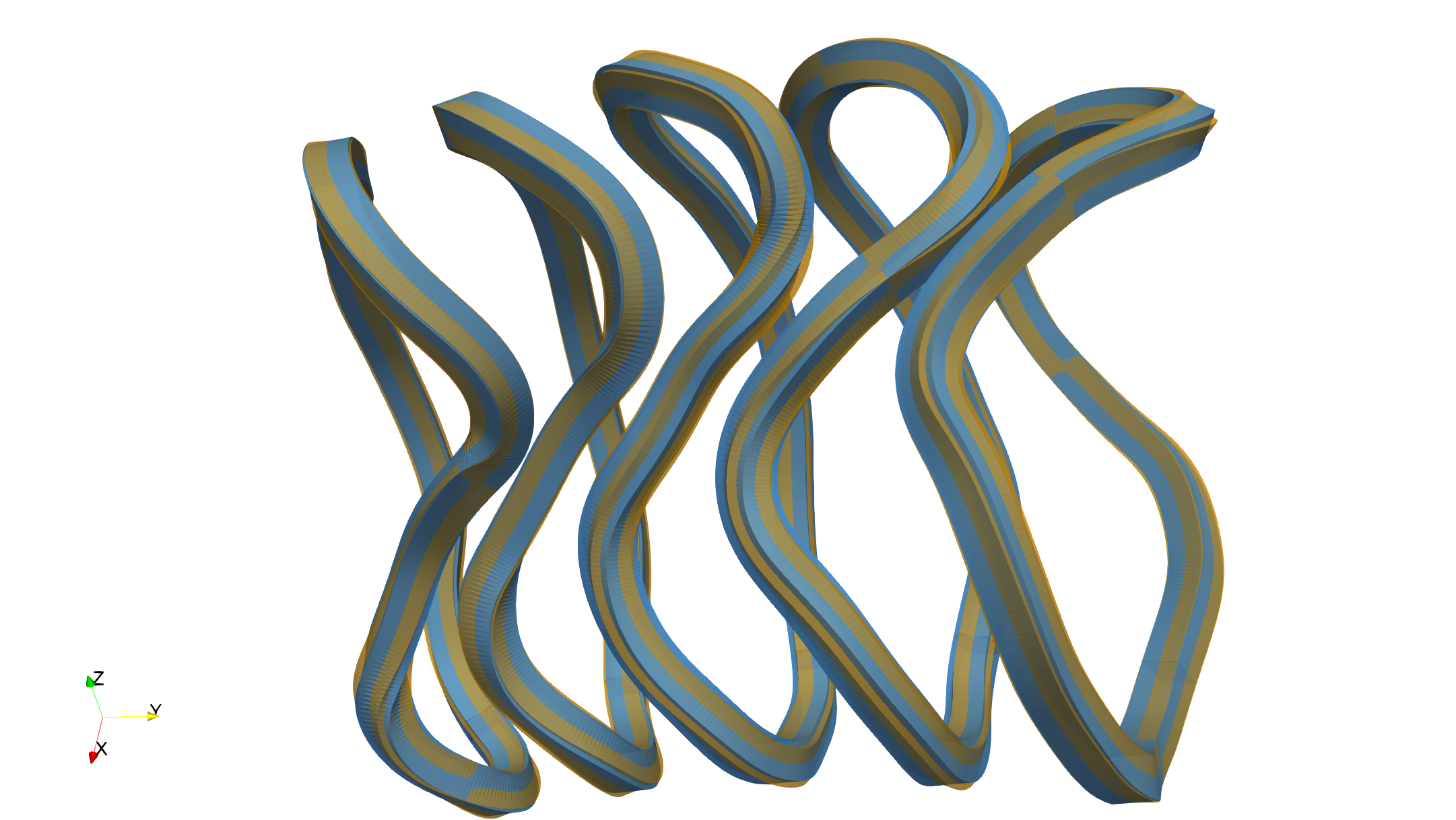}
    \caption{The winding packs (without casing) are displayed for the five independent coils with the pure centroid orientation (blue) and the optimized one (orange).
    While subtle, the difference is visible, especially when focusing on the third coil.}
    \label{fig:reactor_coils_before_after}
\end{figure}

The optimized rotation to be applied on the pure centroid frame is displayed in the left plot of figure \ref{fig:reactor_rotation_angles_and_strain_max} across the coil, for the different coils.
The effect of this rotation is visualized in figure \ref{fig:reactor_coils_before_after}, where the coils with a pure centroid orientation (blue) are compared with the optimized orientation (orange).
The effect is subtle, as the regularization introduced in equation \ref{eq:j_twist} prevents large values of the angles.
Further decreasing the strain in each stack might require a simultaneous optimization of the strain and the central filament of the coil, which is beyond the scope of this study.

Last, we summarize the result of the optimization in the plot on the right of figure \ref{fig:reactor_rotation_angles_and_strain_max}, where we compare the average (circle) and maximum (diamond) strain across all points and all stacks of each coil for the pure centroid (blue) and optimized (orange) orientation.

Thanks to the optimization, all stacks of each coil are below the strain limits with a small but carefully crafted rotation of the winding pack orientation.

\newpage
\section{Summary and Outlook}
\label{sec:summary}
Strain optimization is crucial to ensure the integrity of stellarator coils made from ReBCO superconducting tapes.  In this paper, we implemented a method to optimize the strain on ReBCO tapes in stellarator coils into the stellarator optimization framework \texttt{SIMSOPT}. We penalize hard-way bending and torsion of the tape to lie within the mechanical limits of the superconductor. The winding plane for the tape is constructed from a frame of three mutually orthogonal vectors defined at each quadrature point. The local orientation of the frame and the curve of a filamentary coil are then optimized to minimize strain. 

This method has been implemented into coil optimization and applied to the design of the small stellarators EPOS and CSX as well as a scaled-up, reactor-relevant winding pack for the W7-X stellarator.

The EPOS stellarator aims to confine an electron-positron pair plasma in a quasi-axisymmetric stellarator with ReBCO coils. To inject positrons into the device, we require two larger weave lane coils to create stray field. As a result of a single-stage optimization, we find a suitable coil set with weave lane coils, good quasisymmetry and sufficiently low strain.

CSX will confine a quasiaxisymmetric equilibrium with two interlinked HTS coils. Here, in addition to strain optimization, we avoid net turns of the frame orientation to ease tape winding. Single-stage optimization is performed to obtain a coil set consistent with quasiaxisymmetry and strain constraints without net frame rotation. 

In section \ref{sec:reactor}, the strain optimization approach is applied to the W7-X coils scaled at reactor size, as defined by the ARIES-CS design point.
First, we point out that the variation of curvature and torsion across turns in a reactor-scale winding pack is signficant and needs to be accounted for.
By exploting the orientation of the winding pack as a degree of freedom it is possible to reduce curvature- and torsion-induced strains below limits for every turn of a HTS non-planar coil at reactor scale, while keeping the overall coil shape and the individual orientations of each turn fixed.
Notably, curvature- and torsion-induced strains are not the only sources of HTS strains to consider for a reactor-scale coil.
Other sources, such as the strain generated by the Lorentz load, the manufacturing process of the cable and the assembly of the coil, are outside of the scope of this work and need to be accounted for.

The strain optimization described in this paper can be implemented into objective functions for coil optimization together with other physics and engineering cost fuctions. Future work aims to condense other aspects of superconducting coil engineering, like $\mathbf j \times \mathbf B$ forces and the alignment between the ReBCO crystal axis and the magentic field on the coil into easy-to-compute cost functions. \
\section*{Acknowledgements}
P.H. thanks Jim-Felix Lobsien, Matt Landreman and Dario Panici for useful discussions. This work has been supported by the Helmholtz Association (VH-NG-1430), Simons Foundation Targeted MPS Program (Award 1151685), the PPPL LDRD program, and the Simons Foundation MPS Collaboration Program (Award 560651).. ... [and other grants/etc. from the other authors]. 
\bibliographystyle{jpp}
%
%
%
\bibliography{bibliography_strain_opt_simsopt}
\end{document}